\documentclass[aps,prb,twocolumn,floatfix,superscriptaddress,10pt,showpacs]{revtex4-1}

\usepackage{graphicx}
\usepackage{verbatim}
\usepackage{comment}
\usepackage{amssymb}
\usepackage{enumerate}
\usepackage{psfrag}
\usepackage{pstricks}
\usepackage{color}
\usepackage{hyperref}
\emergencystretch=\maxdimen
\definecolor{ao}{rgb}{0.0, 0.5, 0.0}
\newcommand{\lsim}{\mathrel{\hbox{\rlap{\lower.55ex \hbox{$\sim$}} \kern-.3em \raise.4ex \hbox{$<$}}}}

\begin{document}

\title{Finite-temperature superconducting correlations of the Hubbard model}

\author{Ehsan Khatami}
\affiliation{Department of Physics, University of California, Davis, CA
95616, USA}
\affiliation{Department of Physics and Astronomy, San Jose State
University, San Jose, CA 95192, USA}
\author{Richard T.~Scalettar}
\affiliation{Department of Physics, University of California, Davis, CA
95616, USA}
\author{Rajiv R. P.~Singh}
\affiliation{Department of Physics, University of California, Davis, CA
95616, USA}

\begin{abstract}
We utilize numerical linked-cluster expansions (NLCEs) and the
determinantal quantum Monte Carlo algorithm to study pairing
correlations in the square lattice Hubbard model. To benchmark the NLCE,
we first locate the finite-temperature phase transition of the
attractive model to a superconducting state away from half filling. We
then explore the superconducting properties of the repulsive model for
the $d$-wave and extended $s$-wave pairing symmetries. The pairing
structure factor shows a strong tendency to $d$-wave pairing and peaks
at an interaction strength comparable to the bandwidth.
The extended $s$-wave structure factor and correlation length are larger at
higher temperatures but clearly saturate as temperature is lowered,
whereas the $d$-wave counterparts, which start off lower at high
temperatures, continue to rise near half filling. 
This rise is even more dramatic in the $d$-wave susceptibility.
The convergence of NLCEs breaks down as the susceptibilities and correlation
lengths become large, so we are unable to determine the onset of
long-range order.  However, our results extend the conclusion,
previously restricted to only magnetic and charge correlations, that
NLCEs offer unique window into pairing in the Hubbard model 
at strong coupling.
\end{abstract}


\maketitle

\section{Introduction}

Despite several decades of intensive theoretical research, the question
of whether a non-local attraction can dominate in a fermionic Hubbard
model with local repulsive interaction has remained largely unanswered
for parameters relevant to cuprate high-temperature
superconductors.~\cite{scalapino94,dagotto94,kivelson03,lee06,schrieffer07,scalapino12}
Controlled theoretical approaches confirm this possibility, however, 
 only when the strength of the local repulsion is much smaller than
the hopping amplitude of fermions on a square lattice.~\cite{s_raghu_10}

Numerical methods provide important data for strongly-correlated
quantum Hamiltonians, and, in particular, for phenomena like
superconductivity, magnetism, and Mott metal-insulator transitions.
Although many developments have made these approaches increasingly
powerful over the last decade, significant limitations remain,
especially for fermions.  The density matrix renormalization
group,~\cite{dmrg1,dmrg2} and related techniques, function best in one
dimension. Diagrammatic quantum Monte Carlo
techniques~\cite{n_Prokofev_98,e_kozik_10} are restricted to
weak-coupling regimes. Determinant quantum Monte Carlo
(DQMC),~\cite{dqmc1,dqmc2} and cluster extensions of the dynamic mean-field 
theory~\cite{dca,CDMFT} are limited to real space or momentum space clusters of
tens to hundred of sites. Moreover, the 
``sign problem''~\cite{loh90,iglovikov15} remains an
unsolved problem which limits accessible temperatures unless special
symmetries prevail. 

These limitations emphasize the need for continued algorithm
development.  Recently developed numerical linked-cluster expansions
(NLCEs)~\cite{M_rigol_06,M_rigol_07a,Marcos_07_02,b_tang_12} are
especially promising as an approach to access strong coupling regimes,
which are inaccessible to QMC methods, as a consequence
both of the sign problem and of large and even diverging statistical
fluctuations.  For instance, analysis of magnetic correlations and Mott phases in
trapped atoms on optical lattices,~\cite{r_hart_14,duarte15} where strong
coupling is present at the cloud edge, would not have been possible without NLCEs.

A natural next step is the application of NLCEs to superconductivity.
In this Rapid Communication, we show this method can be developed and
successfully used
to study the pairing correlations in the square lattice Hubbard model, 
\begin{equation}
H=-t\sum_{\left<ij\right> \sigma}c^{\dagger}_{i\sigma}
c^{\phantom{\dagger}}_{j\sigma} + U\sum_i n_{i\uparrow}n_{i\downarrow}
-\mu\sum_{i\sigma} n_{i\sigma},
\label{eq:H}
\end{equation}
where $c^{\phantom{\dagger}}_{i\sigma}$ ($c^{\dagger}_{i\sigma}$)
annihilates (creates) a fermion with spin $\sigma$ on site $i$,
$n_{i\sigma}=c^{\dagger}_{i\sigma} c^{\phantom{\dagger}}_{i\sigma}$ is
the number operator, $U$ is the onsite Coulomb interaction, and $t$ is
the near neighbor hopping integral. We set $k_{\rm B}=1$, and $t=1$ as the 
unit of energy throughout the paper.  

We complement our NLCE results with those obtained from (numerically
unbiased) DQMC simulations on a large lattice. We find excellent
agreement between the two in parameter regions accessible to both, and
show that the lowest temperatures achievable in the NLCE are similar to,
or often lower than, those of the DQMC. For the attractive model
($U<0$), or in the weak-coupling regime of the repulsive model, where
the sign problem is either absent or less severe, DQMC can generally
access lower temperatures than the NLCE. On the other hand, the series
converges to lower temperatures in the strong-coupling regime, where
DQMC runs into sampling difficulties and faces an unforgiving sign
problem. 

We find that, for an interaction strength $U$ equal to the bandwidth,
the $s$-wave pairing structure factor of the attractive model away from
half filling shows divergent behavior at low temperatures, and points to
a finite transition temperature that is consistent with findings of
previous large-scale DQMC 
studies.~\cite{r_scalettar_89,a_moreo_91,t_paiva_04,assaad94,p_staar_14}
For the repulsive model, we consider
several values of $U$ and doping and study pairing in the nonlocal
channels of extended $s$-wave ($s^*$-wave) and $d$-wave. While the structure
factor for the former symmetry tends to saturate at increasingly high
temperatures as the doping is increased, for the latter symmetry, no
such tendency is observed. We examine results at 10\% doping more
closely and find that the low-temperature structure factor is maximum
around $U=8$. On the other hand, the pair-field susceptibility, while larger
for smaller values of $U$ in the intermediate temperature region, shows
a sharp upturn at the lowest accessible temperatures for the largest
interactions considered.

\section{numerical methods}

In NLCEs, an extensive property of the lattice model, when normalized to
the number of sites, is expressed in the thermodynamic limit in terms of
contributions from finite clusters of various sizes and topologies that can
be embedded in the lattice. Thus, NLCEs use the same basis as the
high-temperature expansions (HTEs). However, the calculation of the
extensive quantities at the level of individual clusters is left to an
exact numerical method, such as exact diagonalization, as opposed to a
perturbative expansion in terms of inverse temperature in the HTEs. A
typical expansion involves clusters up to a certain size that are chosen
according to a self-consistent criterion (see below). Despite the lack of an
explicit small parameter, having a finite number of clusters in the
series inevitably leads to the loss of convergence below a certain
temperature, where the correlations in the system extend beyond a length
of the order of the largest sizes considered.  However, the exact
treatment of clusters leads to convergence temperatures that are
 lower than those of HTE with a comparable number of orders. 

Similar to the analytic Pad\'{e} approximations used extensively in
HTEs, here we take advantage of two {\it numerical} resummation
techniques to improve the convergence of our series at low temperatures.
We use the Euler algorithm~\cite{Euler} to resum the last 4-6 terms of
the series or the Wynn algorithm~\cite{Wynn} with 3 and 4 cycles of
improvement (details of these techniques can be found in
Ref.~\onlinecite{M_rigol_07a}). We then take the average of four values,
the last two orders after the Euler and the last two orders after the
Wynn transformations, as our best estimate. To quantify our confidence in
the accuracy of the resummed results, we define a ``confidence region''
around this average where all the values that contribute to the average
fall.  Thus, the errorbars in our figures simply mark the boundaries of
this region and should not be confused with statistical errorbars.

We study the superconducting properties of the model at several values
of the interaction strength and on a fine grid of temperature and
chemical potential. The latter allows us to study the calculated
quantities at constant electronic densities after numerical conversion.
As with previous studies of Hubbard models using the
NLCE,~\cite{Ehsan_Marcos_11,Ehsan_Marcos_12,Tang_12_01} we employ the
site expansion in which the order to which each cluster belongs is
determined by the number of sites it has. In order $l$, we consider all
the open boundary clusters of various shapes and topologies on the
square lattice that have $l$ sites, and use exact diagonalization to
solve for their properties.  For the pairing correlations, the
Hamiltonian matrices are block-diagonalized in each particle number
sector.  So, we are able to carry out the expansion to the ninth order. For
the pairing susceptibility, on the other hand, we can only carry out the
expansion to the seventh order since not only particle number is not
conserved during the time-dependent measurements (see Eq.~\ref{eq:chi}),
but also the majority of the computational time is spent on obtaining
the off-diagonal expectation values, which, like the diagonalization, 
scales like $O(N^3)$.~\cite{note4}
 
DQMC simulations are performed on a $10\times 10$ lattice, which is
large enough to have only small finite size effects at the temperatures
studied here. Results represent averages of at least 8 independent runs
with 10,000 sweeps each. To fix the density, $n$, away from half filling
at each temperature and $U$ value, the chemical potential needs to be
tuned starting from an estimate provided by the NLCE.  Therefore, we
repeat the calculations for several values of $\mu$ to achieve an
accuracy of about 0.01\% for the density. For the structure factor, we
extrapolate our results to the continuous imaginary time limit using the
outcome of two separate simulations with a discretization of the inverse
temperature $\beta=L \Delta \tau$ corresponding to $\Delta\tau=1/16$ and
1/12.  In the case of the susceptibility, 
we choose an even smaller $\Delta\tau=1/50$, in order to perform the
imaginary time integration accurately.  This value leads to Trotter
errors that are negligible in comparison to the statistical ones.

One of the quantities we calculate is the equal-time pairing structure
factor, 
\begin{eqnarray}
S^{\alpha}({\bf q}) = \sum_{\bf r} e^{i {\bf q}\cdot{\bf r}} P^{\alpha}({\bf r}),
\label{eq:pair}
\end{eqnarray}
where 
\begin{equation}
P^{\alpha}({\bf r}_{ij})=\langle \, \Delta^{\alpha\dagger}_i(0) \Delta^{\alpha}_j(0)
+\Delta^{\alpha}_i(0) \Delta^{\alpha\dagger}_j(0)\rangle
\label{eq:p}
\end{equation}
is the equal-time pair-pair correlation function. Here, the pairing operator for the 
symmetry $\alpha$ is defined as
\begin{equation}
\Delta^{\alpha}_i(\tau)= \frac{1}{2}\sum_{j}f^{\alpha}_{ij}
e^{\tau H}(c_{i\uparrow}c_{j\downarrow}-c_{i\downarrow}c_{j\uparrow})e^{-\tau H}.
\end{equation}
We consider three pairing symmetries in this study; (local) $s$-wave,
$d$-wave, and $s^*$-wave. For the $s$-wave symmetry,
$f^{s}_{ij}=\delta_{ij}$. In the case of  $s^*$-wave, $f^{s^*}_{ij}$ is
$+1$ if $i$ and $j$ are nearest neighbors and $j>i$ (to avoid double
counting) and zero otherwise.  $f^{d}_{ij}$ for the $d$-wave symmetry is
the same as $f^{s^*}_{ij}$ except it takes the value $-1$ if the bond
connecting $i$ and $j$ is along the $y$ axis. Here, we consider only the
uniform pairing structure factor, $S^{\alpha}({\bf q}=0)$. The
correlated structure factors, $S^{\alpha}_{\rm corr}$, is obtained by
first subtracting off the uncorrelated parts of the expressions in
Eq.~\ref{eq:p}.\cite{note3}

Having the uniform structure factor ($S^{\alpha}$ or $S^{\alpha}_{\rm corr}$),
the corresponding correlation length, $\xi$, can also be calculated using, e.g.,
\begin{equation}
(\xi^{\alpha}_{\rm corr})^2=\frac{1}{2dS^{\alpha}_{\rm corr}({\bf q}=0)}
\sum_i |{\bf r}_i|^2 P^{\alpha}_{\rm corr}({\bf r}_{0i}),
\end{equation}
where $d=2$ is the dimension.

\begin{figure}[t]
\centerline {\includegraphics*[width=3.3in]{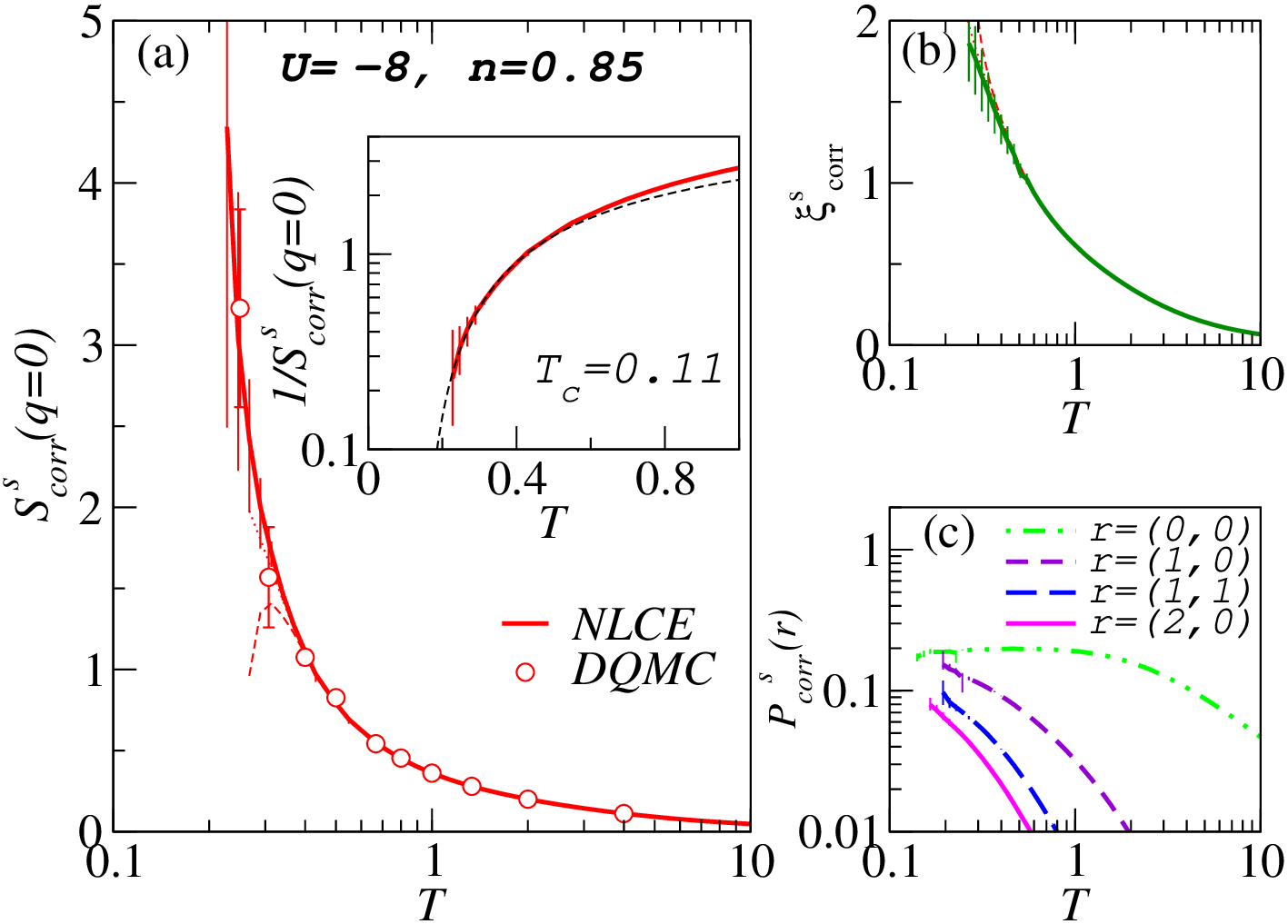}}
\caption{(a) Temperature dependence of the $s$-wave pairing structure
factor at $n=0.85$ for the attractive model with $U=-8$.  The line is
from NLCE and symbols are from DQMC. The inset shows the inverse of the
same function vs $T$ and a low-temperature fit to $A \, {\rm
exp}(B/\sqrt{T-T_c})$ with $T_c=0.11$. (b)-(c) The corresponding
correlation length and short-range correlation functions vs $T$.
In the main panel of (a) and in (b), bare NLCE results before
resummations for the last two orders, $8^{\rm th}$ and $9^{\rm th}$, are
shown as thin dotted and dashed lines, respectively.}
\label{fig:S-8}
\end{figure}

The other quantity of interest for superconductivity is the uniform pairing 
susceptibility, which is defined as
\begin{eqnarray}
\chi^{\alpha} &=& \frac{1}{N}\int_0^\beta \, d \tau\  
\langle \, \mathcal{O}^{\alpha}(\tau) \mathcal{O}^{\alpha\dagger}(0)\rangle,
\label{eq:chi}
\end{eqnarray}
where $\mathcal{O}^{\alpha}(\tau)=\sum_i\Delta^{\alpha}_i(\tau)$.

\section{Results}

We start with the attractive Hubbard model, for which we know there
exists a finite-temperature Kosterlitz-Thouless (KT) phase transition to
an $s$-wave superconducting state away from half
filling.~\cite{r_scalettar_89,a_moreo_91,t_paiva_04,assaad94,p_staar_14}
In Fig.~\ref{fig:S-8}(a), we show the correlated part of the $s$-wave
pairing structure factor from the NLCE for $U=-8$ and at $n=0.85$,
where the superconducting transition temperature is expected to be
maximal.~\cite{t_paiva_04}  Results are in excellent agreement with the 
corresponding DQMC results, plotted as empty circles in that figure. As can 
be inferred from previous DQMC simulations with a smaller $U$,~\cite{t_paiva_04}
finite-size effects in DQMC will not play a role here at temperatures as
low as $T=0.25$.  Whereas the raw NLCE results (before resummations)
converge only to $T\sim0.4$, the averaged value after resummations
suggest a divergent behavior for $S^{s}_{\rm corr}$ at lower
temperatures. They lead us to a regime where we can take advantage of
extrapolations in temperature in order to obtain an estimate for the critical
temperature.  We find that a fit to the KT form [see the inset of
Fig.~\ref{fig:S-8}(a)], leads to $T_c\sim 0.11$, which is in good
agreement with results of past DQMC simulations.~\cite{r_scalettar_89}
The correlation length, which shows an exponential growth, is also
plotted in Fig.~\ref{fig:S-8}(b). Its behavior  is consistent with the
trend seen in Fig~\ref{fig:S-8}(c) for the pairing correlations growing
faster at longer length scales as the temperature is decreased.

\begin{figure}[t]
\centerline {\includegraphics*[width=3.3in]{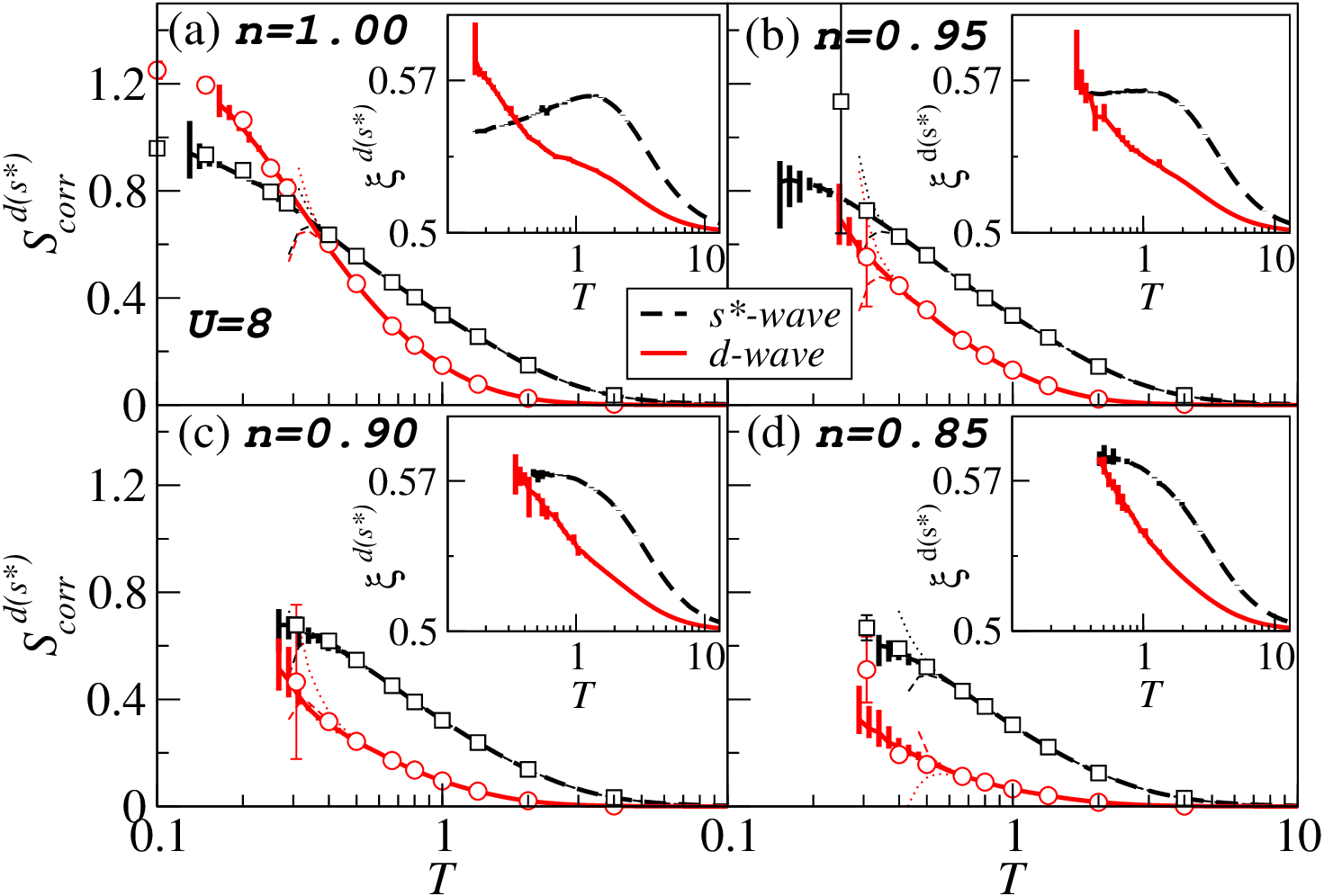}}
\caption{The uniform $d$-wave and extended $s$-wave pairing structure factors
for the repulsive model with $U=8$ at densities 1.00, 0.95, 0.90, and 0.85 
vs temperature. Lines are from the NLCE and symbols 
are from the DQMC. Bare NLCE results before resummations 
for the $8^{\rm th}$ and $9^{\rm th}$ orders, are shown as thin dotted 
and dashed lines, respectively. The insets show the correlation lengths
from the NLCE vs temperature for each case.}
\label{fig:Sdiffn}
\end{figure}

We now turn our focus to the main subject of this study; pairing in the
repulsive Hubbard model. We know that if a similar finite-temperature
transition to a superconducting phase takes place in the latter model,
the pairing symmetry has to be nonlocal because of the onsite
Coulomb repulsion. Therefore, in this case, we only explore the $d$-wave
and the $s^*$-wave symmetries. We also expect the corresponding temperature 
scales to be much smaller than those for the attractive model since we are 
looking for attraction in a repulsive model.

In Fig.~\ref{fig:Sdiffn}, we show the correlated part of the uniform
structure factor for the two pairing symmetries when $U=8$ and at
various average densities. At half filling, the series converges to a
low enough temperature to make clear that $S^{\alpha}_{\rm corr}$
eventually saturates as we decrease the temperature.
In the absence of the `sign problem' at this filling, DQMC can easily 
access lower temperatures. We see in Fig.~\ref{fig:Sdiffn}(a) that, while 
agreeing excellently with NLCE at high temperatures, results from 
DQMC simulations confirm the saturation at lower $T$.
As we move away from half filling into the hole-doped region ($n<1.0$),
an interesting trend is observed; the saturation of the $s^*$-wave
structure factor is seen to take place at higher temperatures whereas
the $d$-wave structure factor continues to grow at the lowest temperatures
accessible to us, although its over all values decrease as we increase
the doping. Hence, if there is an instability to pairing away from half
filling in this model, it would be in the $d$-wave and not the $s^*$-wave
channel. Interestingly, at small dopings near half filling, NLCE results 
are more reliable at generally lower temperatures than those of the 
 DQMC because of the restrictions imposed by a severe sign problem 
in this region [see Fig.~\ref{fig:Sdiffn}(b)]. Nevertheless, results from the 
two methods match within the errorbars at the available temperatures
for all the dopings studied.

\begin{figure}[t]
\centerline {\includegraphics*[width=3.in]{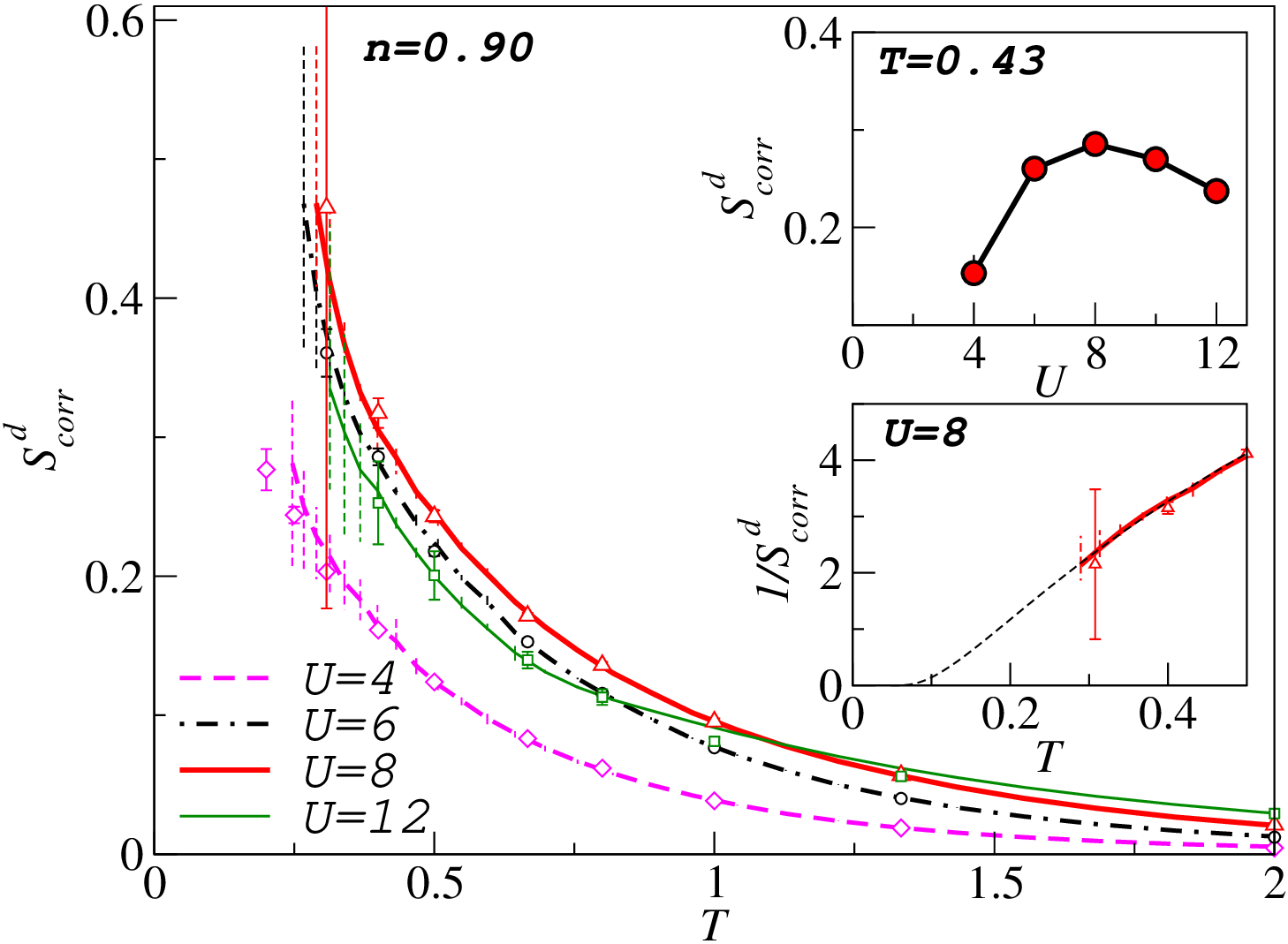}}
\caption{Temperature dependence of the $d$-wave pairing structure factor
at $n=0.90$ for $U=4$, 6, 8, and 12. Symbols are the DQMC results. Top
inset shows the same structure factor vs $U$ at a fixed temperature
$T=0.43$. The bottom inset shows the inverse of the structure factor vs
$T$, along with a fit to the function $A \, {\rm exp}(B/\sqrt{T-T_c})$
for $T<0.6$, which results in $T_c=0.048$.}
\label{fig:SdiffU}
\end{figure}

The favorability of $d$-wave over $s^*$-wave pairing is also evidenced 
by the behavior of the corresponding 
correlation lengths, shown in the insets of Fig.~\ref{fig:Sdiffn}. For
example, even though the low-temperature $s^*$-wave structure factor 
is larger than the $d$-wave one away from half filling, its correlation length
clearly saturates while that of the $d$-wave keeps rising and becomes larger.
The latter can explain the higher convergence temperature of $S^{s^*}_{\rm corr}$ 
in comparison to $S^{d}_{\rm corr}$ in Fig.~\ref{fig:Sdiffn}(b).

Focusing  on $d$-wave pairing at a moderately doped system with
$n=0.9$, we find that at temperatures below one, the structure factor is
largest at $U\sim 8$, which is equal to the non-interacting bandwidth. This can
be seen in Fig.~\ref{fig:SdiffU}, where we show $S^{d}_{\rm corr}$ vs
temperature for $U=4$, 6, 8, and 12. For $U=4$, the DQMC results are
available at lower temperatures than the NLCE and show a relatively slow 
increase of this quantity as the temperature is decreased. In the top inset of 
Fig.~\ref{fig:SdiffU}, we see that the structure factor at $T=0.43$ quickly rises as $U$
increases from 4, reaches a maximum at $U=8$, and then slowly decreases.
Beyond $U=12$, we expect this quantity to scale as $1/U$ as, in the
strong-coupling regime, the only relevant energy scale will be the
exchange interaction of the corresponding low-energy $t-J$ model,
$J=4t^2/U$. The bottom inset in Fig.~\ref{fig:SdiffU} shows the inverse
of the structure factor at $U=8$. Unfortunately, we are not close 
enough to a transition temperature to be able to make any quantitative
statement about its value. However, the best estimate from the DCA for a close 
value of the interaction ($U=7$), puts $T_c$ around 0.05,~\cite{p_staar_14} which is 
consistent with a KT fit to our results for $T<0.6$.

\begin{figure}[t]
\centerline {\includegraphics*[width=3.in]{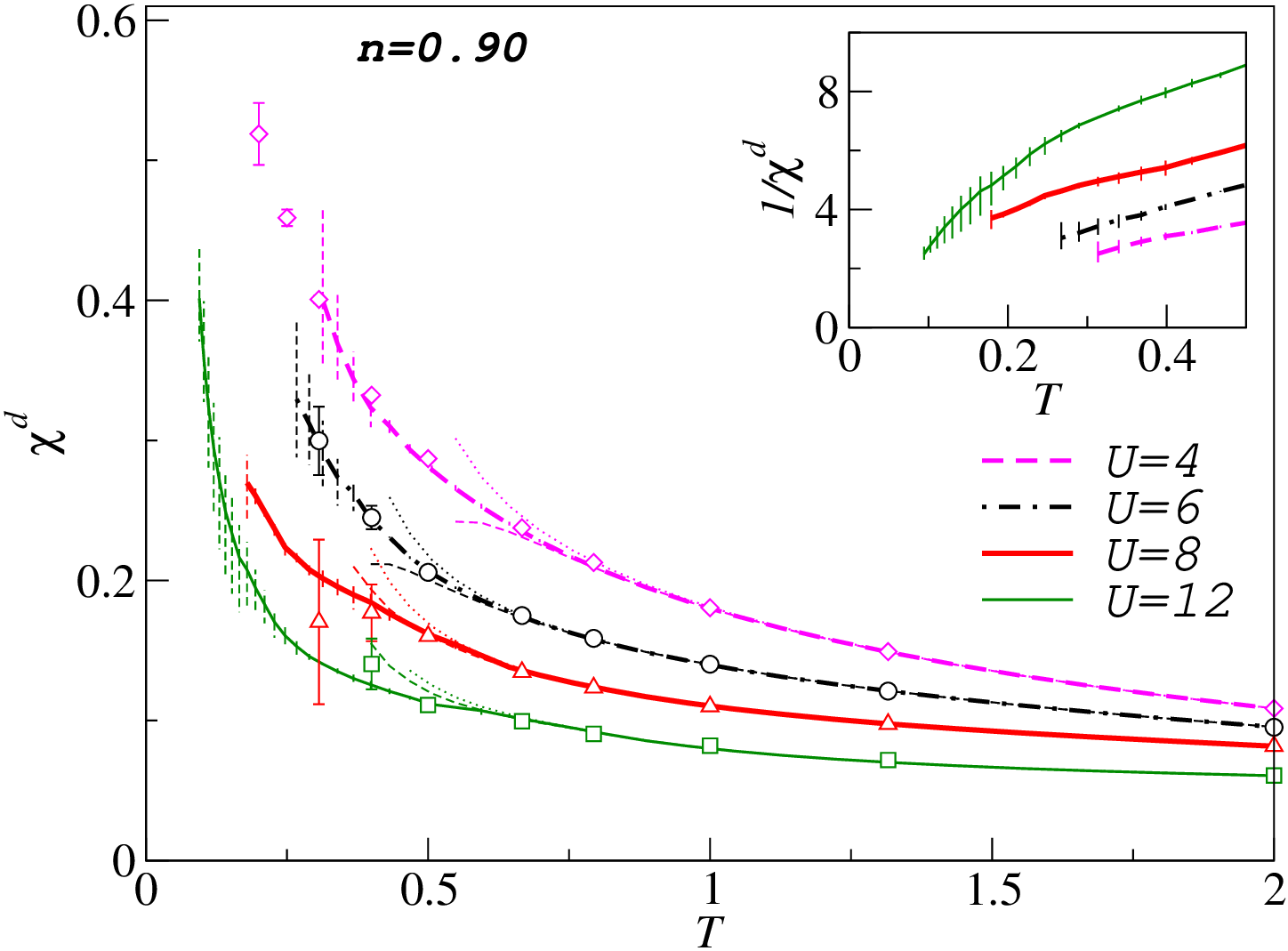}}
\caption{The $d$-wave pairing susceptibility at $n=0.90$ vs temperature for 
$U=4$, 6, 8, and 12. Bare NLCE results before resummations for the last 
two orders, $6^{\rm th}$ and $7^{\rm th}$, are shown as thin dotted and 
dashed lines, respectively. Symbols are the DQMC results. 
The inset shows the inverse of the susceptibility 
vs temperature for the same values of the interaction strength.}
\label{fig:Sus}
\end{figure}

Finally, we turn to the pair-field susceptibility. Figure~\ref{fig:Sus} shows
$\chi^d$ vs temperature at $n=0.9$ for different interaction strengths. 
Our results for the susceptibilities match the DQMC ones very well for smaller $U$ values
and for larger $U$ values when the temperature is not too low. This includes the
susceptibility at $U=4$ and $n=0.875$~\cite{s_white_89} (not shown).
In all cases, there is a rapid increase in
the susceptibility at low temperatures. However, more terms are needed 
for the susceptibility to capture the sharp rise at low temperatures,
and to determine how $T_c$ may depend on $U$.
In future, it would be important to extend the results for the susceptibility to
higher orders and also calculate pairing susceptibilities at non-zero momenta.

In summary, we have employed two unbiased methods, the NLCE and the
DQMC to study finite-temperature superconducting properties of the
square lattice Fermi-Hubbard model. To benchmark our NLCE approach, we first
explore the $s$-wave pairing in the attractive model away from half
filling. By fitting our low-temperature pairing structure factor to
known forms, we obtain a $T_c$ that is consistent with the best estimate
from large-scale QMC simulations. We then investigate the nonlocal
$s^*$-wave and $d$-wave pairing instabilities in the repulsive model at
various dopings and for several interaction strengths. We find that the $d$-wave 
symmetry has the tendency to be dominant at low temperatures and that its structure
factor has a maximum at $U\sim 8$. We also calculate the pairing
susceptibility, which shows a similar divergent behavior in the $d$-wave
channel and a sharp upturn at low temperatures for large interactions.

An important potential application of the results described here is to
ongoing emulation of model Hamiltonians which describe fermionic atoms
in optical lattices. NLCEs allow the rapid evaluation of physical
properties on a dense mesh of Hamiltonian parameters, a requirement for
accurate modeling of optical lattice
experiments~\cite{ole1,ole2,ole3,ole4,r_hart_14,duarte15} where the
confining potential leads to spatially varying chemical potential,
interaction strength, and hopping matrix elements.~\cite{mathey} Here we
have shown the potential importance of NLCEs as a tool to analyze
pairing in these systems.

\acknowledgements

This work was supported by the Department of Energy under DE-NA0001842-0
(EK and RTS), and by the National Science Foundation (NSF) grant number
DMR-1306048 (EK and RRPS). This work used the Extreme Science and
Engineering Discovery Environment (XSEDE) under project number
TG-DMR130143, which is supported by NSF grant number OCI-1053575.


\begin{thebibliography}{41}
\bibitem [{not({\natexlab{a}})}]{note4}%
  \BibitemOpen
  \href@noop {} {} \bibinfo {note} {In the ninth order
  alone, we need to distinguish all 1285 clusters that may have the same
  topology (the same Hamiltonian matrix), but are not related by point group
  symmetry. The largest matrices to be diagonalized have a linear size of
  $N=15876$. In the seventh order, there are 108 symmetrically distinct
  clusters. The largest matrices to be diagonalized here have a linear size of
  $N=3432$.}\BibitemShut {Stop}%
\end{thebibliography}
\end{document}